\documentclass[12pt]{article}
\usepackage{graphicx}
\usepackage{epstopdf}
\usepackage{caption}
\usepackage{subcaption}
\usepackage{subfig}

\usepackage{epsfig,amsmath,amssymb,verbatim,mathrsfs}

\def\beq{\begin{eqnarray}}
\def\eeq{\end{eqnarray}}
\def\bea{\begin{eqnarray}}
\def\eea{\end{eqnarray}}

\newcommand{\rf}{R_\mathcal{F}}

\newcommand{\be}{\begin{equation}}
\newcommand{\ee}{\end{equation}}

\begin{document}

\setlength{\baselineskip}{0.2in}


\begin{titlepage}
\noindent
\flushright{March 2013}
\vspace{0.2cm}

\begin{center}
  \begin{Large}
    \begin{bf}
Minimal Tree-Level Seesaws with a Heavy Intermediate Fermion\\

     \end{bf}
  \end{Large}
\end{center}

\vspace{0.2cm}

\begin{center}

\begin{large}
{Kristian~L.~McDonald}\\
     \end{large}
\vspace{0.5cm}
  \begin{it}
ARC Centre of Excellence for Particle Physics at the Terascale,\\
School of Physics, The University of Sydney, NSW 2006, Australia\\\vspace{0.5cm}
\vspace{0.1cm}
klmcd@physics.usyd.edu.au
\end{it}
\vspace{0.5cm}

\end{center}


\begin{abstract}
There exists a generic minimal tree-level diagram, with two external scalars and a heavy intermediate fermion, that can generate naturally small neutrino masses via a seesaw. This diagram has a mass insertion on the internal fermion line, and the set of such diagrams can be partitioned according to whether the mass insertion is of the Majorana or Dirac type. We show that, once subjected to the demands of naturalness (i.e. precluding small scalar vacuum expectation values that require fine-tuning), this set is finite, and contains a relatively small number of elements. Some of the corresponding models have appeared in the literature. We present the remaining original models, thus generalizing the Type-I and Type-III seesaws, and apparently exhausting the list of their minimal non-tuned variants.

\end{abstract}

\vspace{1cm}

\end{titlepage}

\setcounter{page}{1}


\vfill\eject


\section{Introduction\label{sec:introduction}}
\begin{figure}[ttt]
\begin{center}
        \includegraphics[width = 0.5\textwidth]{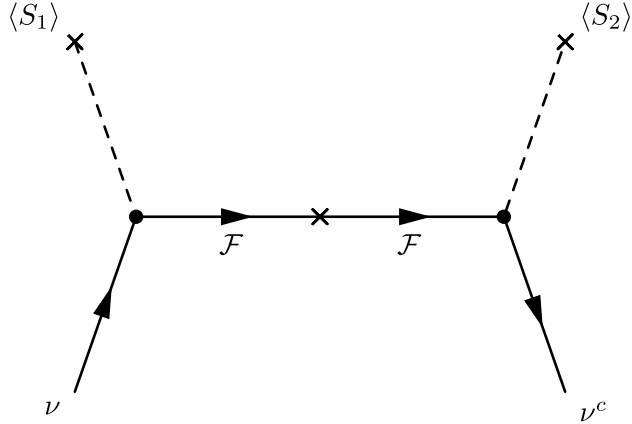}
\end{center}
\caption{Generic tree-level diagram for small neutrino mass from heavy fermion exchange. }\label{fig:nu_tree_generic}
\end{figure}

There exists a generic minimal tree-level diagram, with two external scalars and a heavy intermediate fermion, that can generate  naturally suppressed Standard Model (SM) neutrino masses; see Figure~\ref{fig:nu_tree_generic}. The internal fermion line in this diagram has a single mass insertion, which can be of the Majorana type or the Dirac type. The minimal (and best known)  models that produce this diagram are the Type-I~\cite{type1_seesaw} and Type-III~\cite{Foot:1988aq} seesaws, where the SM is augmented by an $SU(2)_L$ singlet/triplet fermion with zero hypercharge. In these cases, lepton number symmetry is broken by the (Majorana) mass insertion. However, the underlying mechanism is more general, and alternative extensions of the SM can realize the basic diagram in a number of ways.

The set of these minimal tree-level diagrams can be partitioned according to the nature of the mass insertion (equivalently, to the origin of lepton number violation).\footnote{The partition is formally defined by introducing a relation, such that two diagrams are related when the internal mass insertion is of the same type, namely Majorana or Dirac. This relation has the necessary properties (reflexive, symmetric and transitive) to define an equivalence relation and thus uniquely partitions the set.} Any given representative diagram of the distinct subsets has a number of properties that can be determined without recourse to a specific model. These properties can, in turn, be used to guide  one in the search for viable realizations of these minimal seesaws. 

In this work we aim to catalogue the minimal models that produce small neutrino masses via one of these tree-level diagrams. To achieve this goal we  reverse-engineer the models. In the process we ``rediscover" some models that have already appeared in the literature, and discover a number of additional models which, to the best of our knowledge, have not previously appeared.

As one moves beyond the minimal realizations of Figure~\ref{fig:nu_tree_generic}, the models typically require additional fields to be added to the SM. Thus, it naively appears that the tree-level diagram can be realized in a large number of ways. However, as we will see, if one restricts their attention to natural models, in which no tuning is needed to achieve small vacuum expectation values (VEVs), the list of candidate models is finite and quite short. We provide a comprehensive version of this list, apparently exhausting the variant seesaws of this type. Interestingly a number of the corresponding models only realize viable seesaws when the new physics occurs near the TeV scale, and are therefore largely amenable to discovery (or exclusion) at the LHC. These models realize neutrino mass by low-energy effective operators with mass dimension $d>5$.

Before proceeding we note that some models described in this paper employ scalars in non-fundamental representations of $SU(2)_L$. The demands of perturbative unitarity place general upper bounds on the quantum numbers of larger scalar multiplets~\cite{Hally:2012pu} (also see~\cite{Earl:2013jsa}). However, all multiplets appearing here  are consistent with these constraints. Bounds from flavor changing processes in models with large multiplets can also be found in Ref.~\cite{Liao:2010rx}.

The plan of this paper is as follows. In Section~\ref{sec:mass_insert} we consider diagrams with a lepton number violating Majorana mass insertion.  Section~\ref{sec:L_vertex}  considers diagrams with a Dirac mass insertion, for which lepton number symmetry is broken by a vertex. Loop effects are briefly discussed in Section~\ref{sec:loops}, and we conclude in Section~\ref{sec:conc}. In an Appendix we provide details for some non-minimal cases with a Majorana mass insertion. The ``busy" reader is advised that our main results are contained in the two tables presented in the text. Readers interested primarily in the particle content of the models can refer to these tables; the minimal natural models with a mass insertion of the Majorana (Dirac) type are the first three (five) entries in Table~\ref{L_mass_result} (Table~\ref{L_vertex_result}).
\section{Models with a Majorana Mass Insertion\label{sec:mass_insert}}
\begin{figure}[ttt]
\begin{center}
        \includegraphics[width = 0.5\textwidth]{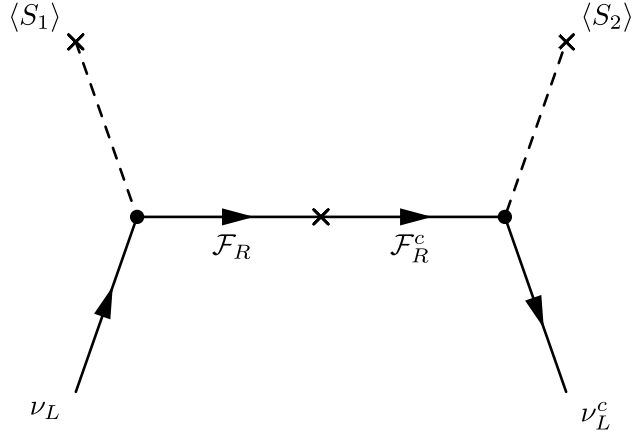}
\end{center}
\caption{Generic tree-level diagram for small neutrino mass from heavy fermion exchange when lepton-number symmetry is broken by the (Majorana) mass insertion.  The simplest realizations are the Type-I and Type-III seesaws, with $\mathcal{F}_R\sim(1,1,0)$ and $\mathcal{F}_R\sim(1,3,0)$, respectively, and with $S_1=S_2=H\sim(1,2,1)$.}\label{fig:L_massinsert_nu_tree_generic}
\end{figure}

In the generic tree-level diagram of Figure~\ref{fig:nu_tree_generic}, the mass insertion can be of the Majorana type or the Dirac type. Adopting the standard convention for a Type-I seesaw, in which the beyond-SM fermion $\mathcal{F}$ is assigned the same lepton number value as the SM leptons, these two cases correspond to lepton number violation by the mass insertion, or by a vertex, respectively. In this section we consider models with a lepton number violating mass insertion, for which the generic tree-level diagram takes the form shown in Figure~\ref{fig:L_massinsert_nu_tree_generic}. Inspection of the figure reveals the following generic features:
\begin{itemize}
\item The internal fermion transforms as $\mathcal{F}_R\sim (1,R_\mathcal{F},0)$ under the SM gauge symmetry; that is, $\mathcal{F}_R$ should form a real representation of $SU(3)_c\times SU(2)_L\times U(1)_Y$.
\item The multiplet $\mathcal{F}_{R}$ should contain an electrically neutral component, constraining $R_{\mathcal{F}}$ to be an odd number.
\item The two external scalars can be distinct. However, minimal models occur when they have the same quantum numbers: $S_1=S_2\sim(1,R_S,Y_S)$. 
\item The quantum numbers of the scalars must satisfy the following conditions:
\bea
Y_S=-Y_L=1 \qquad\mathrm{and}\qquad R_S\otimes R_{\mathcal{F}}\supset2.
\eea
\end{itemize} 
We focus on the minimal case with $S_1=S_2$ here, but comment on the more general case at the end of the section. 

The first condition ensures that a lepton-number violating Majorana mass appears in the Lagrangian, while the last condition ensures that the requisite Yukawa couplings appear:
\bea
\mathcal{L}&\supset& \lambda_S \,\overline{\mathcal{F}_R}\;S_1\, L + \frac{M_{\mathcal{F}}}{2}\;\overline{\mathcal{F}_R}\;\mathcal{F}_R^c+\mathrm{H.c.}\;,
\eea
where $L\sim(1,2,-1)$ denotes a SM lepton doublet. Integrating out the heavy fermions, and inserting the scalar VEVs, gives the generic form for the seesaw-suppressed SM neutrino masses in these models:
\bea
m_\nu \simeq \lambda_S^2\,\frac{\langle S_1\rangle ^2}{M_{\mathcal{F}}}.
\eea 
This mass has the familiar seesaw form, and for good reason; the simplest models that realize Figure~\ref{fig:L_massinsert_nu_tree_generic} occur when $S_1=S_2=H\sim(1,2,1)$ is the SM scalar. In this case, one can have $\mathcal{F}_R\sim(1,1,0)$ or $\mathcal{F}_R\sim(1,3,0)$, corresponding to the well known Type-I~\cite{type1_seesaw} and Type-III~\cite{Foot:1988aq} seesaw mechanisms, respectively. These give rise to the famous $d=5$ Weinberg operator~\cite{Weinberg:1979sa} in the low-energy effective theory (see e.g.~\cite{Ma:1998dn}).

It should be clear that the basic mechanism for generating seesaw suppressed neutrino masses is more generic than the Type-I and Type-III seesaws. Indeed, on the basis of the above-mentioned conditions, it seems that one can select any odd value of $R_{\mathcal{F}}$, and then choose an appropriate $S_1$ to obtain suppressed neutrino masses via Figure~\ref{fig:L_massinsert_nu_tree_generic}. However, there are additional considerations when one moves away from the simplest cases. 

In models with $S_1\ne H$, one has $\rf\geq 5$ and $R_S>2$, so $S_1$ forms a non-trivial representation of $SU(2)_L$. Thus, when symmetry breaking occurs,  the VEV of $S_1$ contributes to the $W$ and $Z$ boson masses. This modifies the tree-level value of the $\rho$-parameter away from the SM value of $\rho=1$, leading to a bound on the VEV of roughly $\langle S_1\rangle\lesssim \mathcal{O}(1)$~GeV~\cite{Nakamura:2010zzi}. Direct searches, on the other hand, typically constrain the mass of scalars with non-trivial $SU(2)_L$ quantum numbers  to satisfy $M_{S}\gtrsim\mathcal{O}(100)$~GeV. Therefore $S_1$ is expected to be a heavy scalar with a  small VEV: $\langle S_1\rangle/M_S\ll1$.

There are two ways in which one can add a heavy scalar to the SM and obtain a small tree-level VEV. In the first approach, one must tune the parameters in the scalar potential, to ensure that multiple dimensionful quantities that are $\gtrsim\mathcal{O}(100)$~GeV conspire to deliver ($\langle S_1\rangle/\mathrm{GeV})\lesssim\mathcal{O}(1)$. This approach has limited appeal. 

In the second approach, the quantum numbers of $S_1$ are such that the scalar potential contains a term that is linear in $S_1$. We denote this term as $-\mu^3 S_1\subset V(H,S_1)$, where $\mu$ denotes generic dimensionful quantities (field operators or constants). Then $S_1$ develops a nonzero VEV of the form $\langle S_1\rangle=\mu^3/M_S^2$.  Thus, provided $M_S\gg \mu$, the VEV $\langle S_1\rangle$ can be naturally suppressed, ensuring that the constraints from the $\rho$-parameter are satisfied without fine tuning.\footnote{This VEV suppression is analogous to what happens in the Type-II seesaw~\cite{type2_seesaw}, where the scalar potential has the term $-\mu HTH\subset V(H,T)$ and the triplet $T\sim(1,3,2)$ obtains the VEV $\langle T\rangle = \mu\langle H\rangle^2/M_T^2$. We restrict our attention to models possessing a generalized version of this natural VEV suppression.}

The second approach is clearly more desirable. We therefore restrict our attention to models in which the small VEV of any new scalar is naturally suppressed without fine-tuning. These models achieve a \emph{natural} tree-level seesaw via heavy fermion exchange (with a Majorana mass insertion, in the present case).

To find candidate models that meet this constraint, one can consider the direct product of $n$ SM scalars, namely $H^n\otimes$. Here $H$ denotes a generic SM scalar, including hermitian- or charge-conjugate fields, and the integer $n$ satisfies $n\leq3$. A viable model is found if the direct product contains a term $(1,R_n,-1)\subset H^n\otimes $ for some $n\leq3$, provided a scalar field with the quantum numbers $(1,R_n,1)$ contains an electrically neutral component. This fixes the quantum numbers of the new scalar, $S_1\sim(1,R_n,1)$, and determines a new model with a natural tree-level seesaw. 

The demand of naturalness, in the sense defined above, turns out to be a rather stringent constraint. A comprehensive study of the aforementioned direct products reveals a single candidate; namely 
\bea
\tilde{H}\otimes H^\dagger\otimes H\supset (1,4,-1),
\eea
where $\tilde{H}$ is the charge conjugate field. Thus, if we add the scalar field $S_1\sim(1,4,1)$ to the SM, and also add a real fermion $\mathcal{F}_R$ with quantum numbers $\mathcal{F}_R\sim(1,\rf,0)$ such that $\rf\otimes 4 \supset 2$, we obtain a natural tree-level seesaw. There are two candidates for the new fermion field, namely $\mathcal{F}_R\sim(1,3,0)$ and  $\mathcal{F}_R\sim(1,5,0)$. The former case is non-minimal as the standard Type-III seesaw is  also present. Absent hierarchical Yukawa couplings of a conspiratorial nature, the Type-III term is expected to dominate the neutrino mass matrix (we will return to this matter below). 

We  therefore conclude that one can extend the SM  by adding the fields
\bea
S_1\sim(1,4,1) \quad\mathrm{and}\quad\mathcal{F}_R\sim(1,5,0),
\eea
to arrive at a new seesaw model that is distinct from the standard Type-I and Type-III seesaws. Actually, this model was recently proposed in Ref.~\cite{Kumericki:2012bh}, where it was shown to provide a viable model of naturally suppressed seesaw neutrino masses. The Lagrangian contains the terms
\bea
\mathcal{L}&\supset&
 \lambda_S \;\overline{\mathcal{F}_R}\;S_1\, L + \frac{M_{\mathcal{F}}}{2}\;\overline{\mathcal{F}_R}\;\mathcal{F}_R^c +\lambda\,H^3\, S_1\, ,
\eea
where the last term ensures that $S_1$ develops a naturally suppressed VEV:\footnote{The full potential is discussed in Ref.~\cite{Kumericki:2012bh}; similar to the Type-II seesaw, provided the (positive) scalar mass satisfies $M_S^2\gg\langle H\rangle^2$, the linear term in $S$ forces a naturally suppressed VEV of this form.}
\bea
\langle S_1\rangle\simeq \lambda\, \frac{\langle H\rangle^3}{M_S^2}.\label{s1_vev_quad}
\eea
Putting all the pieces together, one arrives at the final expression for the seesaw suppressed neutrino masses:
\bea
m_\nu&\simeq& \lambda_S^2\,\frac{\langle S_1\rangle ^2}{M_{\mathcal{F}}}\ \sim \  \frac{\langle H\rangle ^6}{M_{\mathcal{F}}\,M_S^4}\,,\label{eq:quintuplet_nu_mass}
\eea
where, for simplicity,  we consider dimensionless couplings of $\mathcal{O}(1)$ in the last expression. The first expression has the standard seesaw form, $m_\nu \sim\langle S_1\rangle^2/M_{\mathcal{F}}$, bearing testament to the relation between this model and the more familiar seesaws. Denoting $M_S\sim M_{\mathcal{F}}\equiv M$ as an approximate common scale for the  new physics, the second expression gives $m_\nu\sim\langle H\rangle ^6/M^5$. Thus, in this model, the tree-level diagram in Figure~\ref{fig:L_massinsert_nu_tree_generic} produces an effective low-energy operator\footnote{The uniqueness of neutrino mass operators with $d>5$ is studied in Ref.~\cite{Liao:2010ku}.} with mass-dimension $d=9$, namely: $\mathcal{O}_\nu=L^2H^6/M^5$. The new fields can therefore be relatively light; dimensionless couplings of order $\mathcal{O}(0.1)$ allow $M$ to be as low as $\mathcal{O}(100)$~GeV~\cite{Kumericki:2012bh}. The scale of new physics is thus expected to be significantly smaller than that found in the minimal realizations of Figure~\ref{fig:L_massinsert_nu_tree_generic}, namely the Type-I and Type-III seesaws. The payoff for the extra complexity, it seems, is that interesting regions of parameter space for the resulting model can be explored at the LHC, and there is a realistic chance that the full parameter space of the model can be probed in the ``not-too-distant" future.

This exhausts the list of minimal models that produce natural seesaw neutrino masses via Figure~\ref{fig:L_massinsert_nu_tree_generic}. The natural models of this type are the Type-I and Type-III seesaws, and the model with a quintuplet fermion introduced in Ref.~\cite{Kumericki:2012bh}. These models are minimal, in the sense that they comprise minimal extensions of the SM, and that there is only one type of tree-level diagram generating neutrino masses. There are additional models with a lepton number violating mass insertion that are less minimal. These models generate diagrams with $S_1\ne S_2$. In what follows we briefly discuss an example to illustrate some of the differences.  Additional general analysis for these models appears in Appendix~\ref{app:mass_non_minimal}.

Return to the case where the SM is extended to include the scalar $S_1\sim(1,4,1)$. Instead of considering the minimal scenario with $\mathcal{F}_R\sim(1,5,0)$, let us now take $\mathcal{F}_R\sim(1,3,0)$. Then there are three distinct tree-level diagrams that contribute to light neutrino masses. The first of these has two external SM scalars, the second has two external scalars $S_1$, and the third has one external $S_1$ and one external SM scalar. The light neutrino mass has multiple contributions, which can be written as
\bea
m_\nu &\simeq&\lambda_H^2\,\frac{\langle H\rangle^2}{M_{\mathcal{F}}} \ +\  2\lambda_H\,\lambda_S \,\frac{\langle S_1\rangle\langle H\rangle}{M_{\mathcal{F}}}\ +\  \lambda_S^2 \,\frac{\langle S_1\rangle^2}{M_{\mathcal{F}}}\,.\label{eq:type_3_variant}
\eea
The first piece here is the usual Type-III seesaw term. If all dimensionless couplings are of a similar order of magnitude, this Type-III term is expected to dominate the ``pure $S_1$" term and the mixing term, due to the relation $\langle H\rangle\gg\langle S_1\rangle$; see the Appendix.  Another way to say this, is that the three terms in Eq.~\eqref{eq:type_3_variant} correspond to low-energy operators with mass dimension $d=5$, $d=7$ and $d=9$, respectively. Given that all three operators arise from tree-level diagrams, one would normally expect the operator with the lowest mass dimension to dominate. The model is therefore a more complicated version of the Type-III seesaw, with additional subdominant contributions to the neutrino mass matrix.

Strictly speaking, one can imagine that for some reason the coupling $\lambda_H$ is much smaller than the coupling $\lambda_S$, so the hierarchy in VEVs is overcome. In this case the mixing term and the Type-III term can be subdominant to the pure $S_1$ piece, giving $m_\nu\sim\langle S_1\rangle^2/M_{\mathcal{F}}$, with additional subdominant Type-III (and mixed) contributions. This region of parameter space produces a seesaw that is distinct from the aforementioned minimal variants, and the model can be studied as a viable theory of neutrino masses. Of course, the need for hierarchical Yukawa couplings runs counter to the spirit of the seesaw mechanism. However, the model does have a certain benefit, as the phenomenology of the new scalar can be distinct, and there is a better chance of the new physics being experimentally accessible than in the pure Type-III seesaw~\cite{Ren:2011mh}.


\begin{table}
\centering
\begin{tabular}{|c|c|c|c|c|c|}\hline
& & & & &\\
\ \ Model\ \ &
$S_1$ & $\mathcal{F}_R$ &
$S_2$&$\ \ \ [\mathcal{O}_\nu]\ \ \ $&Ref.\\
& & & & &\\
\hline
& & & & &\\
$(a)$& $(1,2,1)$ &$(1,1,0)$ &$-$&$d=5$ &\ Type-I seesaw\ \\ 
& & & & &\\
\hline
& & & & &\\
$(b)$& $(1,2,1)$ &$(1,3,0)$ &$-$&$d=5$ &\ Type-III seesaw\ \\ 
& & & & &\\
\hline
& & & & &\\
$(c)$& $(1,4,1)$ &$(1,5,0)$ &$-$&$d=9$ &\cite{Kumericki:2012bh}\\ 
& & & & &\\
\hline
& & & & &\\
$(d)$& $(1,2,1)$ &$(1,3,0)$ &$(1,4,1)$&$d=5,7,9$ &\cite{Ren:2011mh}\\ 
& & & & &\\
\hline
& & & & &\\
$(e)$& $(1,4,1)$ &$(1,5,0)$ &$(1,6,1)$&$\ d=9,11,13\ $ &?\\ 
& & & & &\\
\hline
\end{tabular}
\caption{\label{L_mass_result} Natural Seesaw Models with a Majorana Mass Insertion.  The first three entries are minimal while last two entries are not; model $(d)/(e)$ is essentially model $(b)/(c)$ with an additional field.}
\end{table}


There is one  other natural non-minimal model, which is obtained by adding the field\footnote{The in-built mechanism that triggers a naturally suppressed VEV for $S_2\sim(1,6,1)$ is discussed in the Appendix; related discussion appears in the next section.} $S_2\sim(1,6,1)$  to the model with $S_1\sim(1,4,1)$ and $\mathcal{F}_R\sim(1,5,0)$. As with the preceding example,  absent parameter hierarchies the neutrino mass matrix is expected to be dominated by the $S_1$-term found in Eq.~\eqref{eq:quintuplet_nu_mass}. However, Yukawa coupling hierarchies can allow the ``pure $S_2$" piece to dominate. This second example exhausts the natural non-minimal models of this type, as we explain in Appendix~\ref{app:mass_non_minimal}.

To conclude this section, we note that we have considered the generic tree-level seesaw with two external scalars and a heavy intermediate fermion, in the case where lepton number symmetry is broken by a Majorana mass insertion. Naively it appears that many variants of this seesaw are possible. However,  if one restricts their attention to natural models, in which small VEVs are obtained without fine-tuning, it appears that only five models are possible. The minimal models are the Type-I and Type-III seesaws, and the model  proposed recently in Ref.~\cite{Kumericki:2012bh}, which employs a real quintuplet of fermions. The two non-minimal natural models are the variant of the Type-III seesaw discussed in Ref.~\cite{Ren:2011mh}, and the variant of the quintuplet fermion model described in the previous paragraph. We summarize these results in Table~\ref{L_mass_result}, where the question mark indicates that, to the best of our knowledge, the model is original.

\section{Models with a Dirac Mass Insertion\label{sec:L_vertex}}
In addition to the models of the preceding section, which contain a Majorana mass insertion, the class of seesaws described by Figure~\ref{fig:nu_tree_generic} includes cases with a Dirac mass insertion. In these models lepton number symmetry is broken at one of the vertices. We consider such models in this section.

\begin{figure}[ttt]
\begin{center}
        \includegraphics[width = 0.5\textwidth]{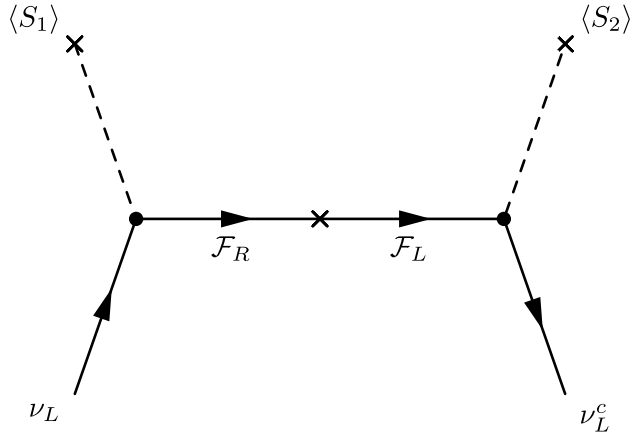}
\end{center}
\caption{Generic tree-level diagram for small neutrino mass when the mass insertion is of the Dirac type.  In the simplest realization, one of the scalars is the SM doublet, $S_1=H\sim(1,2,1)$, and the quantum numbers for the new fields are uniquely determined, with $\mathcal{F}_{L,R}\sim(1,3,-2)$, and $S_2\sim(1,4,-3)$.}\label{fig:L_vertex_nu_tree_generic}
\end{figure}

The generic diagram  for models with a Dirac mass insertion is given in Figure~\ref{fig:L_vertex_nu_tree_generic}. The figure reveals the following generic features for these models:
\begin{itemize}
\item The intermediate fermion is vector-like: $\mathcal{F}_{L,R}\sim(1,\rf,Y_\mathcal{F})$.
\item The scalars $S_{1,2}$ are necessarily distinct, with quantum numbers $S_{1,2}\sim(1,R_{1,2},Y_{1,2})$.
\item Consideration of the Yukawa couplings leads to the following relations:
\bea 
2\,Y_L&=& Y_1\,+\,Y_2\,,\nonumber\\
\rf&\subset& R_L\otimes R_1\, ,\nonumber\\
\rf&\subset& R_L\otimes R_2\,,
\eea
where $R_L=2$ is the SM value for the lepton doublet $L$.
\item As a result of the above relations, and the fact that $R_H=R_L$, one deduces the following:
\bea
(R_H\otimes R_1)\otimes(R_H\otimes R_2)\supset 1\,.
\eea
\end{itemize}
The last point is important; when combined with the hypercharge relations, it tells that, in addition to the Yukawa couplings needed to generate the seesaw in Figure~\ref{fig:L_vertex_nu_tree_generic}, the model \emph{automatically} contains a renormalizable term of the form $H^2S_1S_2$ in the scalar potential. Thus, we know the Lagrangian contains the following terms
\bea
\mathcal{L}&\supset& \lambda_1 \,\overline{L}\,S_1\, \mathcal{F}_R\,+\, \lambda_2\,\overline{L}\,S_2\,\mathcal{F}_L^c \,+\,M_{\mathcal{F}}\,\overline{\mathcal{F}_L}\mathcal{F}_R\,+\,\lambda\, H^2S_1S_2.\label{eq:lagrange_L_vertex}
\eea
Evaluating the leading order seesaw contribution to the neutrino mass matrix, one finds
\bea
m_\nu&\simeq& \lambda_1\lambda_2\,\frac{\langle S_1\rangle \langle S_2\rangle}{M_{\mathcal{F}}}.
\eea
Modulo the dependence on the two distinct VEVs, this is of the standard seesaw form.

The comments made thus far are completely generic; they apply to all models of this type, without any consideration given to the specific quantum numbers of the fields $S_{1,2}$ and $\mathcal{F}$. Ideally, we would like to now specify the viable realizations of this type of tree-level seesaw. 

The most obvious case to consider is the minimal one, obtained by choosing $S_1=H\sim(1,2,1)$. This model turns out to be uniquely specified and, in addition to the fermion $\mathcal{F}$, only one new (beyond SM) scalar is required. The quantum numbers for the fermion are fixed by the term $\overline{L}H\mathcal{F}_R$ to be $\mathcal{F}\sim(1,3,-2)$, which then fixes the quantum numbers of the other scalar as $S_2\sim(1,4,-3)$. This model was recently proposed in Ref.~\cite{Babu:2009aq}, while various phenomenological aspects of the fermion triplet $\mathcal{F}\sim(1,3,-2)$ were studied in Ref.~\cite{DelNobile:2009st}. Note that $\langle S_2\rangle\ne0$ contributes to electroweak symmetry breaking and, given that $R_2>2$, this VEV is constrained by $\rho$-parameter measurements: it  must satisfy $\langle S_2\rangle\lesssim\mathcal{O}(1)$~GeV. Thus, in this model one requires a heavy new scalar with a small VEV. However, the model has an inbuilt feature which renders this demand harmless, in the sense that no fine tuning is required. The last term in Eq.~\eqref{eq:lagrange_L_vertex} now becomes $S_1S_2H^2\rightarrow H^3S_2$. This term ensures that $S_2$ develops an induced VEV once electroweak symmetry breaking is triggered by $\langle H\rangle\ne0$. When the scalar $S_2$ is heavy the VEV takes the form~\cite{Babu:2009aq}:
\bea
\langle S_1\rangle&\simeq& \lambda\, \frac{\langle H\rangle^3}{M_S^2},
\eea
which is naturally suppressed by $M_S\gg \langle H\rangle$. Denoting a common scale for the new physics as $M_S\sim M_\mathcal{F}\equiv M$, one has $m_\nu\sim \langle H\rangle^4/M^3$, and neutrino masses arise from the $d=7$ operator $\mathcal{O}_\nu=L^2H^4/M^3$. The model is therefore completely natural and provides a viable alternative to the more traditional Type-I and Type-III seesaws.

Moving away from this minimal model, the more general case has both $S_1$ and $S_2$ as beyond SM fields. Naively, it again appears that the list of viable combinations for $S_{1,2}$ and $\mathcal{F}$ is large. However, all new scalars in such models form $SU(2)_L$ representations with $R_{1,2}>2$, and the VEVs $\langle S_{1,2}\rangle$ are therefore subject to the constraints from the $\rho$-parameter.\footnote{There is an exception to this statement; there exists beyond-SM scalars whose quantum numbers ensure that $\rho=1$ at tree-level, the smallest such example being the septuplet $S\sim(1,7,4)$. We find, however, that the use of this field to generate  a minimal tree-level seesaw is not viable due to a generalized version of the accidental global $U(1)$ symmetry discussed in Ref.~\cite{Hisano:2013sn}. When the suptuplet develops a VEV a Goldstone boson thus arises, contrary to observations. One can presumably add more fields to explicitly break this symmetry; given our focus on minimal models we do not pursue this matter.} Thus, we expect such models to require heavy new scalars with small VEVs. 

As with the analysis in Section~\ref{sec:mass_insert}, we can limit the scope of our study by restricting our attention to \emph{natural} models, in which the small VEVs arise without any fine-tuning in the scalar potential. We can then ask ``What are the necessary conditions for naturally small VEVs to occur?" Eq.~\eqref{eq:lagrange_L_vertex} shows that the scalar potential in these models always contains a term that is linear in both $S_1$ and $S_2$ (the ``$\lambda$-term"). Indeed, in the minimal case with $S_1=H$, this term is precisely responsible for ensuring that $S_2$ acquires a naturally suppressed VEV. A consideration of the general scalar potential shows that this term is not sufficient to ensure that both VEVs $\langle S_{1,2}\rangle$ are naturally suppressed in the general case. However, provided one of the fields, $S_i$, obtains a naturally suppressed VEV due to terms in the potential $V(H,S_i)\subset V(H,S_1,S_2)$ ($i=1,2$), the $\lambda$-term ensures that $S_{j}$  also acquires a naturally suppressed VEV:
\bea
\langle S_j\rangle &\simeq & \lambda \,\frac{\langle S_i\rangle \langle H\rangle^2}{M_j^2}\,\quad\mathrm{for}\quad i\ne j\,.\label{eq:small_vev_vertex}
\eea
This generalizes the result found in Ref.~\cite{Babu:2009aq}, and is a generic feature of models that contain the tree-level diagram with a Dirac mass insertion shown in Figure~\ref{fig:L_vertex_nu_tree_generic}. Note that if, in a particular model, the potential $V(H,S_i)\subset V(H,S_1,S_2)$ contains a term linear in $S_i$ for both $i=1$ and $i=2$, a small value of $\langle S_i\rangle\ne0$ can be generated for each $S_i$ via the interactions with $H$ in $V(H,S_i)$. However, due to Eq.~\eqref{eq:small_vev_vertex} only one scalar is \emph{required} to have a linear term in $V(H,S_i)$ to arrive at a natural model.

The strategy for determining the general set of natural models is now apparent. One is after a scalar $S_i$ with quantum numbers $S_i\sim(1,R_i,Y_i)$, such that the potential $V(H,S_i)$ contains a term that is linear in $S_i$. This ensures that $S_i$ has a naturally suppressed VEV, and Eq.~\eqref{eq:small_vev_vertex} then ensures a naturally small VEV for $S_j$. Once a suitable scalar $S_i$ is found, the viable sets of quantum numbers for $\mathcal{F}$ and $S_j$ are determined by the Yukawa couplings. One must therefore consider the direct product of $n$ SM scalars, namely $H^n\otimes$, where $H$ denotes a generic SM scalar, including hermitian- or charge-conjugate fields, and the integer $n$ satisfies $n\leq3$. A viable model is found if the direct product contains a term $H^n\otimes\supset (1,R_i,-Y_i)$ for some $n\leq3$, provided a scalar field with the quantum numbers $(1,R_i,Y_i)$ contains an electrically neutral component. 

As in Section~\ref{sec:mass_insert}, the list of viable candidates for $S_i$ is rather short, though the fact that the hypercharge value is not fixed to be $Y=1$ provides some additional freedom in the present case. We find the following candidates:
\begin{itemize}
\item$H^\dagger\otimes \tilde{H}\supset (1,3,-2)$,
\item $H^\dagger\otimes H\supset (1,3,0)$,
\item $\tilde{H}\otimes H^\dagger\otimes H\supset (1,4,-1)$,
\item
$\tilde{H}\otimes H^\dagger\otimes \tilde{H}\supset (1,4,-3),$
\end{itemize}
The first candidate can be excluded because  the triplet $T\sim(1,3,2)$ leads to a Type-II seesaw, which is expected to dominate the diagram with an intermediate fermion. For this reason we exclude any models that require a triplet $T$.\footnote{We do not include the singlet $(1,1,0)$ in the above list precisely because it requires the second scalar to be the triplet $T$.} The remaining three cases can be systematically explored to determine the viable models.

To see how the above information is used to obtain complete models, consider the second case on the list, which gives $S_1\sim(1,3,0)$. Inspection of the Yukawa coupling $\overline{L}S_1\mathcal{F}_R$ reveals that either $\mathcal{F}_R\sim(1,2,-1)$ or $\mathcal{F}_R\sim(1,4,-1)$. However, the first case requires $S_2=T$, which we exclude, so that only $\mathcal{F}_R\sim(1,4,-1)$ is viable. The second Yukawa coupling  requires either $S_2=T$ or $S_2\sim(1,5,-2)$. Excluding the first case, we arrive at a new model with a natural tree-level seesaw by extending the SM to include the fields $S_1\sim(1,3,0)$, $\mathcal{F}_R\sim(1,4,-1)$ and $S_2\sim(1,5,-2)$. To the best of our knowledge this is an original seesaw model.\footnote{The phenomenology of the real triplet was studied in Ref.~\cite{FileviezPerez:2008bj} (and see references therein).}

\begin{figure}[ttt]
\begin{center}
        \includegraphics[width = 0.5\textwidth]{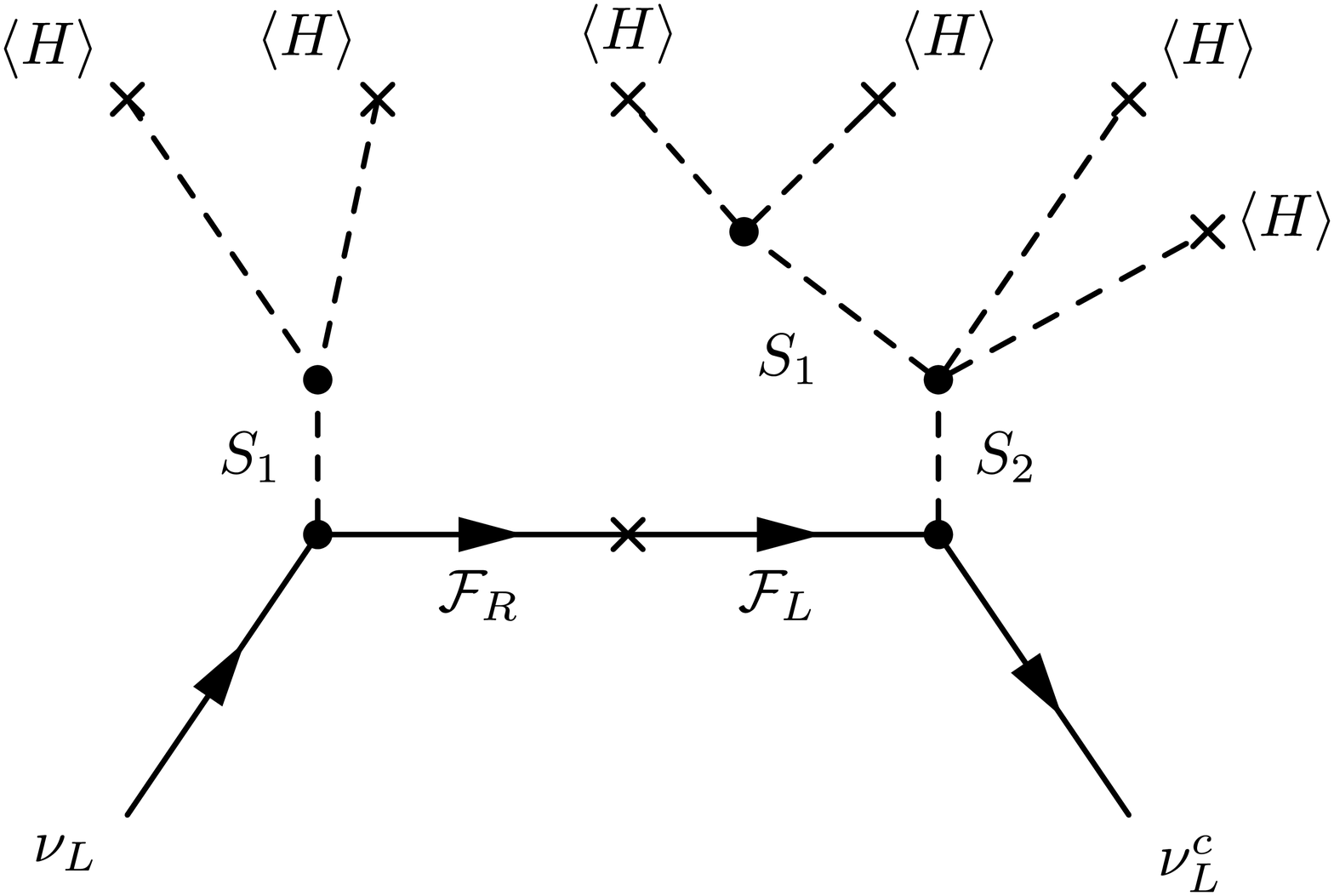}
\end{center}
\caption{A new tree-level seesaw involving the real scalar triplet $S_1\sim(1,3,0)$, the fermion quadruplet $\mathcal{F}\sim(1,4,-1)$, and the scalar quintuplet $S_2\sim(1,5,-2)$.}\label{fig:d9_triplet_nu_tree}
\end{figure}

Let us consider neutrino masses in this model. By construction, the potential $V(H,S_1)\subset V(H,S_1,S_2)$ contains the term $-\mu H^\dagger S_1 H$, which induces a naturally suppressed VEV for the real triplet:
\bea
\langle S_1\rangle& \simeq &\mu\, \frac{\langle H\rangle^2}{M_1^2}.
\eea
As should be clear from the above discussion, the full scalar potential automatically contains the requisite terms to induce a suppressed VEV for $S_2$, giving
\bea
\langle S_2\rangle& \simeq &\lambda\,\, \frac{\langle S_1\rangle \langle H\rangle^2}{M_2^2}\, .
\eea
The seesaw suppressed neutrino mass is
\bea
m_\nu&\sim &  \frac{\langle S_1\rangle \langle S_2\rangle}{M_{\mathcal{F}}}\ \sim\ \frac{\langle S_1\rangle^2 \langle H\rangle^2}{M_{\mathcal{F}}\,M_2^2}\ \sim\  \frac{\mu^2}{M_{\mathcal{F}}}\, \frac{\langle H\rangle^6}{M_1^4\,M_2^2}.
\eea
The first form for $m_\nu$ demonstrates the relation to the more familiar seesaw expressions. If the dimensionful parameters $M_{\mathcal{F}}$, $M_{1,2}$, and $\mu$ are denoted by an approximate common scale $M$, the final expression gives $m_\nu\sim \langle H\rangle^6/M^5$. Neutrino masses are therefore produced by an effective operator in the low-energy theory with mass dimension $d=9$, namely $\mathcal{O}_\nu=L^2H^6/M^5$. It is clear that the new fields can have masses of $\mathcal{O}(\mathrm{TeV})$, and we again see the payoff for the additional complexity in these models; though less minimal, they have the advantage of being experimentally testable. For completeness we present the tree-level diagram showing the $d=9$ nature of this model in Figure~\ref{fig:d9_triplet_nu_tree}.


This example shows how realistic seesaw models are constructed with the above information. We will not present the process for the other cases but instead simply list the results in Table~\ref{L_vertex_result}. Of these six models, model (A) is the aforementioned minimal variant~\cite{Babu:2009aq}, and model (C) was presented in Ref.~\cite{Picek:2009is}. The four remaining models are, to the best of our knowledge, presented here for the first time. Note that one of these models~[(F)] is non-minimal, as neutrino mass is generated by two distinct tree-level diagrams, though the particle content is no more elaborate. This model gives rise to a $d=9$ diagram due to the particle content listed but also generates the tree-level diagram of model (A).\footnote{Additional non-minimal models can be obtained by combining the particle content in models (C) and (D), or models (D) and (E). We do not include these in the table as they are ``more non-minimal," requiring more than three distinct beyond-SM multiplets.}

\begin{table}
\centering
\begin{tabular}{|c|c|c|c|c|c|}\hline
& & & & &\\
\ \ Model\ \ &
$S_1$ & $\mathcal{F}_R$ &
$S_2$&$\ \ [\mathcal{O}_\nu]\ \ $&\ \ Ref.\ \ \\
& & & & &\\
\hline
& & & & &\\
$(A)$& $\ (1,2,1)\ $ &$\ (1,3,-2)\ $ &$\ (1,4,-3)\ $&$d=7$ &\cite{Babu:2009aq} \\ 
& & & & &\\
\hline
& & & & &\\
$(B)$& $(1,3,0)$ &$(1,4,-1)$ &$(1,5,-2)$&$d=9$ &?\\ 
& & & & &\\
\hline
& & & & &\\
$(C)$& $(1,4,1)$ &$(1,5,-2)$ &$(1,4,-3)$&$d=9$ &\cite{Picek:2009is}\\ 
& & & & &\\
\hline
& & & & &\\
$(D)$& $(1,6,1)$ &$(1,5,-2)$ &$(1,4,-3)$&$d=11$ &?\\ 
& & & & &\\
\hline
& & & & &\\
$(E)$& $(1,4,1)$ &$(1,5,-2)$ &$(1,6,-3)$&$d=11$ &?\\ 
& & & & &\\
\hline
& & & & &\\
$(F)$& $(1,4,1)$ &$(1,3,-2)$ &$(1,4,-3)$&$\ d=7,9\ $ &?\\ 
& & & & &\\
\hline
\end{tabular}
\caption{\label{L_vertex_result} Natural Seesaw Models with a Dirac Mass Insertion. The first five entries are minimal while last one is not; model $(F)$ is essentially model $(A)$ with an additional field.}
\end{table}

\section{Comments on Loops\label{sec:loops}}
\begin{figure}[t]
\begin{center}
        \includegraphics[width = 0.5\textwidth]{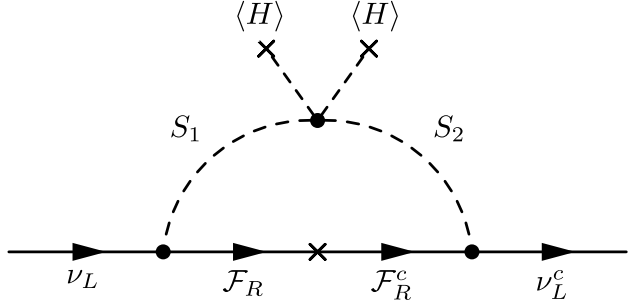}
\end{center}
\caption{A loop diagram obtained by joining the two external scalars in the tree-level diagram with a Majorana mass insertion.}\label{fig:loop}
\end{figure}

Before concluding, we note that, in addition to the tree-level diagrams in the models we have described, there will also be loop processes that contribute to the light neutrino masses. Generally speaking, one expects loop effects to be subdominant to the tree level results. However, one must be careful to account for the details of the given model. In the present class of models, there always exists a coupling $\lambda S_1S_2H^2\subset V(H,S_1,S_2)$, where $S_{1,2}$ can denote the SM scalar. Consequently one expects that the two external scalars ($S_{1,2}$) can be joined together to form a one-loop diagram with two external SM scalars; see Figure~\ref{fig:loop}. This one-loop diagram generates an effective operator in the low-energy theory with mass dimension $d=5$, i.e.~the Weinberg operator.  Many of the tree-level diagrams presented here produce low-energy operators with mass dimension $d>5$. For these models there will be regions of parameter space in which the tree-level diagram is not the dominant effect. 

For example, Ref.~\cite{Kumericki:2012bh} finds regions of parameter space in which the loop-diagram dominates the tree-diagram for model (c) in Table~\ref{L_mass_result}. One expects that similar effects are possible in the models presented here when $[\mathcal{O}_\nu]\ge 7$. In general both tree-level and loop effects will be  present, with the loop-effects dominating for larger values of the beyond-SM masses. If the loop-effect dominates the tree-level mass in a given model, one obtains radiative neutrino masses with sub-dominant tree-level corrections. Our interest is in tree-level seesaw masses and one can crudely estimate the parameter space for which the tree-level diagram is dominant as follows. In a model with $[\mathcal{O}_\nu]=d$, the tree-level mass is roughly $m_{\mathrm{tree}}\sim \langle H\rangle^{d-3}/M^{d-4}$, while the loop-mass is $m_{\mathrm{loop}}\sim (1/16\pi^2) \langle H\rangle^2/M$. Demanding that the former exceeds the latter gives $M \lesssim(16\pi^2)^{\frac{1}{d-5}} \langle H\rangle$, so that $M\lesssim\{ 2,\; 0.6,\; 0.4\}$~TeV for $d=\{7,9,11\}$.  The region of parameter space for which the tree-level mass dominates is therefore eminently testable  for models with $d=9$. While tree-level dominance in models with $d=11$ is, at best, marginally permitted, and may already be excluded. Certainly the LHC should definitively determine whether the tree-level masses are viable for models (D) and (E), and likely for models (B) and (C) also. Meaningful regions of parameter space will also be probed for model (A). Let us emphasize, however, that this rough estimate should only be used as a guide.

\section{Conclusion\label{sec:conc}}
There is a generic minimal tree-level diagram, with two external scalars and a heavy intermediate fermion, that can naturally  achieve small neutrino masses via a seesaw. The diagram has a mass insertion on the internal fermion line, and the set of such diagrams can be partitioned according to whether the mass insertion is of the Majorana or Dirac type.  We have shown that, upon demanding that requisite small VEVs are naturally suppressed, this set is finite, and amenable to a systematic description. We have undertaken this task, and found a number of original seesaw models in the process.  

Our results show that there are only three minimal models with a lepton-number symmetry breaking (Majorana) mass insertion; namely the Type-I and Type-III seesaws, and the quintuplet seesaw of Ref.~\cite{Kumericki:2012bh}. Five additional minimal models also exist, in which the mass insertion is of the Dirac type, and lepton number symmetry is broken by a vertex; these are the $d=7$ model of Ref.~\cite{Babu:2009aq}, two models with $d=9$ (one of which was presented in Ref.~\cite{Picek:2009is}), and two further models with $d=11$.  Our work appears to exhaust the list of these minimal non-tuned tree-level seesaws. 

In a partner paper we shall present similar generalizations of the inverse seesaw mechanism~\cite{Law:2013gma}. A more detailed analysis of the new $d=9$ model, and a discussion of dark matter candidates in the class of models considered here~\cite{Law:2013saa}, will appear elsewhere.
\section*{Acknowledgments\label{sec:ackn}}
The author thanks S.~S.~C.~Law. This work was supported by the Australian Research Council.
\appendix
\section{Next-to-Minimal Models with a Majorana Mass Insertion\label{app:mass_non_minimal}}
Models that generate neutrino mass by the seesaw diagram in Figure~\ref{fig:L_massinsert_nu_tree_generic} with $S_1\ne S_2$ also generate diagrams with two external $S_1$ VEVs and two external $S_2$ VEVs. These models are therefore non-minimal, in the sense that there are distinct tree-level diagrams contributing to the neutrino masses. One can always remove one of the scalars from the model and yet retain a viable theory of seesaw neutrino masses. 

In these models the intermediate fermion must be real, $\mathcal{F}_R\sim(1,\rf,0)$, with odd $\rf$. The quantum numbers of the scalars are fixed as $S_{1,2}\sim(1,\rf\mp1,1)$. Thus,  to ensure that $R_{1}=\rf-1 \ge2$, only models with $\rf\ge3$ can have two scalars with distinct quantum numbers. For these models one always has $(1,\rf,0)\subset  \tilde{H}\otimes S_1$ and $(1,\rf,0)\subset  \tilde{H}\otimes S_2$, so the term $\lambda\tilde{H}^2S_1S_2\subset V(H,S_1,S_2)$ is always allowed. Provided one of the scalars obtains a naturally suppressed VEV via its couplings with the SM scalar $H$ in $V(H,S_i)\subset V(H,S_1,S_2)$, the other scalar $S_j$ can acquire a naturally suppressed VEV via this quartic coupling.

For the case of $\rf=3$ one has the model of Ref.~\cite{Ren:2011mh}, while for $\rf=5$ one has $S_{1,2}\sim(1,5\mp1,1)$, and $S_1$ has suitable quantum numbers to allow $\lambda S_1H^3\subset V(H,S_2)$. Both $S_{1}$ and $S_2$ therefore develop naturally suppressed VEVs. For $\rf>5$ no coupling linear in $S_i\sim(1,\rf\mp1,1)$ can appear in $V(H,S_i)$, so one cannot be assured of generating naturally suppressed VEVs for the beyond-SM scalars [which form larger representations of $SU(2)_L$]. The list of natural non-minimal  models of this type is therefore exhausted by the last two entries in Table~\ref{L_mass_result}.

\end{document}